# Magnetization plateau and anisotropic magnetoresistance in the frustrated Kondo-lattice compound Ce$_3$ScBi$_5$

Zhongchen Xu, [1,2] Yaxian Wang, [3] Cuiwei Zhang, [2,3] Hongxiong Liu, [2,3] Xin Han, [2,3] Liusuo Wu, [4] Xianmin Zhang, [1*] Quansheng Wu, [3,5*] and Youguo Shi, [2,3,5*]

[1] *Key Laboratory for Anisotropy and Texture of Materials (Ministry of Education), School of Material Science and Engineering, Northeastern University, Shenyang 110819, China*

[2] *Center of Materials Science and Optoelectronics Engineering, University of Chinese Academy of Sciences, Beijing 100190, China*

[3] *Beijing National Laboratory for Condensed Matter Physics and Institute of Physics, Chinese Academy of Sciences, Beijing 100190, China*

[4] *Department of Physics, Southern University of Science and Technology, Shenzhen 518055, China*

[5] *Songshan Lake Materials Laboratory, Dongguan, Guangdong 523808, China*

To whom correspondence should be addressed.
*ygshi@iphy.ac.cn
*quansheng.wu@iphy.ac.cn
*zhangxm@atm.neu.edu.cn

# Abstract

Kondo metals with geometric frustration offer fertile ground for exploring exotic states of matter with a field-induced fractional magnetization platform and nonsaturating magnetoresistance. Herein, a $Ce_3ScBi_5$ single crystal with anti-$Hf_5Sn_3Cu$ hexagonal structure was successfully synthesized via the bismuth self-flux method, leading to the formation of cerium cations arranged in a frustrated structure within a distorted kagome lattice. Magnetic measurements exhibit two distinct antiferromagnetic transitions at 4.1 and 5.9 K. Specifically, the occurrence of multiple metamagnetic transitions between magnetization plateaus is evidenced upon applying magnetic fields perpendicular to the $c$ axis. Transport measurements highlight remarkable Kondo-lattice characteristics and anisotropic magnetoresistance in $Ce_3ScBi_5$. The anomalous Hall contributions are observed at low temperatures under critical fields, suggesting Fermi surface reconstruction in a subset of the metamagnetic transitions. We have constructed a temperature-field phase diagram to provide comprehensive information on the complex magnetic structures arising from competitive interactions. Our work establishes $Ce_3ScBi_5$ and related materials as a unique platform for exploring low-dimensional quantum fluctuations in bulk crystals, and analyzes the critical role of geometric frustration in Kondo and Ruderman-Kittel-Kasuya-Yosida physical frameworks.

## II. INTRODUCTION

Geometric frustration arises from the competition of exchange coupling between localized spins, which represents a crucial research direction for future investigations into quantum magnetic systems. Quantum fluctuations resulting from different degeneracies can impede long-range ordering and give rise to intriguing physical phenomena, such as quantum spin liquids, magnetization platforms and spin ice[1-3]. On the other hand, the Kondo-lattice belongs to the classical strongly correlated electron system, which enables the realization of the quantum phase transition from long-range magnetic order to a nonmagnetic heavy fermion liquid phase[4]. The competition between the Ruderman-Kittel-Kasuya-Yosida (RKKY) indirect exchange and the Kondo effect influences the coupling between local 4f electrons and itinerant electrons, ultimately determining the ground state properties of the Kondo lattice. By applying a magnetic field or pressure, the Kondo-lattice exhibits peculiar behavior characterized by unconventional superconductivity in $CeCoIn_5$ and quantum criticality in $YbRh_2Si_2$ at extremely low temperatures[5, 6].

Frustration as a new regulatory dimension has recently been added to classic Kondo physics and has aroused great interest among researchers[7-9]. The quantum fluctuation induced by the frustration will impact the competition between the Kondo effect and the RKKY interaction. For example, the distorted kagome-lattice YbAgGe exhibits intricate magnetic states under applied magnetic fields and displays non-Fermi liquid behavior[10]. The rare-earth triangular-lattice $TmMgGaO_4$ serves as an experimental platform for investigating the two-dimensional random transverse-field Ising model[11, 12]. These findings underscore the urgent need for discovering and synthesizing new single-crystal frustrated Kondo lattices in order to advance physics research.

Based on the aforementioned motivation, our focus was directed towards the $Ln_3MX_5$ family of compounds[13-16] (Ln = rare-earth; M = transition materials; X = As, Sb, Bi), wherein the arrangement of rare-earth elements forms a distorted kagome structure. Previous studies[17-19] have confirmed significant Kondo lattice features in Cerium-based 315 compounds. The study demonstrates the quasi-one-dimensional properties of $La_3MnAs_5$[20] and $La_3CrAs_5$[21], revealing that the complex magnetic mechanism is influenced by interchain interactions between one-dimensional (1D) $MAs_6$ spin chains. Furthermore, $Sm_3ZrBi_5$[22] emerges as a topological candidate due to its contribution from quasi-one-dimensional Bi chains. However, research on frustration within the $Ln_3MX_5$ system remains insufficient, and further investigation is required.

In this study, a novel Ce-based intermetallic compound, $Ce_3ScBi_5$, has been unveiled as a member of the $Ln_3MX_5$ family, showcasing intriguing frustrated Kondo-lattice behavior, with the advantage of further insight into the competing low-energy states in quasi-1D lattices. A meticulous analysis of magnetic orders and magnetization plateaus sheds light on the impact of spin frustration and reveals a complex magnetic ground state present in $Ce_3ScBi_5$. Specific heat and transport measurements demonstrate remarkable Kondo lattice characteristics. Particularly striking is the observation of a large anisotropy in magnetoresistance within this material. Additionally, anomalous Hall contributions are observed at low temperatures under critical fields. For the rich and peculiar physical properties of $Ce_3ScBi_5$, we highlight the significance of its magnetization platform and the impact of anisotropic magnetoresistance in strongly correlated systems.

## II. EXPERIMENTAL DETAILS

Single crystals of $Ce_3ScBi_5$ were grown using the bismuth self-flux method. Elemental reagents, including Ce (lump, 99.9%, Purui Advanced Material Technology Co. Ltd., Beijing, China), Sc (piece, 99.9%, Alfa Aesar), and Bi (pill, 99.999%, Alfa Aesar), were combined in a ratio of 1:4:10 and loaded into an alumina crucible. The crucible was then sealed in a quartz tube under approximately 0.1 atm of argon gas pressure. The tubes were heated at a rate of 100 K/h to 1323 K. Samples were allowed to be homogenized at 1323 K for 18 h before being cooled to 973 K at a rate of 1.5 K/h, and then centrifuged to separate the crystals from the excess Sc-Bi flux. Hexagonal rodlike morphologies, as shown in Figure 1(b), were obtained on the bottom of the crucible. For measurements, we recommend using a scalpel to remove residual Bi flux on the surface of single crystals. However, it should be noted that prolonged exposure to air for several hours would cause decomposition.

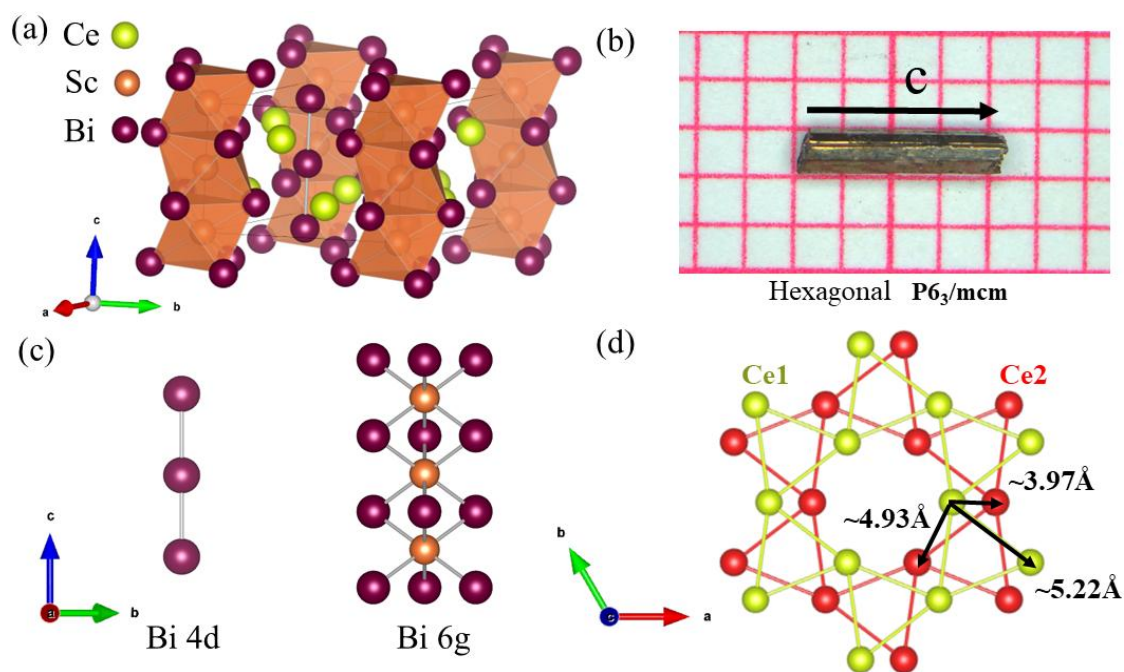

Figure 1. Structure of $Ce_3ScBi_5$. (a) Overall crystal structure. (b) Single crystal photograph on a 1-mm grid paper, highlighting the hexagonal rod-like geometry. (c) Main view of the 1D Bi chain (left) and the 1D $ScBi_6$ face-sharing octahedral (right) motifs. (d) Top view of the Ce-based distorted Kagome lattice along the c axis.

## III. RESULTS AND DISCUSSION

*Crystal structure.* $Ce_3ScBi_5$ is a new member of the $Ln_3MX_5$ family, with an anti-$Hf_5Sn_3Cu$ hexagonal structure of the space group $P6_3/mcm$ (No. 193), as depicted in Figure 1(a). The ($h00$) peaks are very sharp, indicating excellent crystalline quality (Fig. S3(a) in the Supplemental Material[23]). The crystal structure of $Ce_3ScBi_5$ has three primary lattices to focus on: the 1D Bi chain, the 1D $ScBi_6$ face-sharing octahedral, and the Ce-based distorted kagome lattice. The Bi atoms in Figure 1(c) occupy two distinct Wyckoff positions: Bi1 atoms located at the 4d site (2/3,1/3,1/2), forming a 1D Bi chain parallel to the c-direction, while Bi2 atoms reside at the 6g site (1, y, 3/4), positioned at the vertex of the shared octahedron. The Bi−Bi bond length within the 1D Bi chain is measured to be slightly longer at 3.2590(5) Å compared to classical Zintl-Klemm bonds [3.225(5) Å][24, 25]. The elongated Bi−Bi bonds can be considered hypervalent [26]. Due to the relative absence of significant s-p interactions, hypervalent Bi ions in 1D linear chains have a formal electron count of seven and adopt the –2 state under a half-filled $p_z$ orbital[27, 28].

Figure 1(d) illustrates that the arrangement of magnetic Ce atoms in the *ab* plane constitutes a frustrated structure of a distorted kagome lattice, alternatively stacking with the *AB*-type sequence along the c-axis. The nearest-neighbor distance between Ce atoms in Kagome layers is approximately 5.22 Å, which is slightly larger than the distance between Ce atoms in adjacent layers (~4.93 and 3.97 Å). The spin-exchange interaction between adjacent layers plays an indispensable role in determining the magnetic structure[29, 30]. Given the RKKY interaction and Kondo effect between $Ce^{3+}$ cations on a distorted kagome lattice, it is intriguing to investigate the magnetic ordering of $Ce_3ScBi_5$.

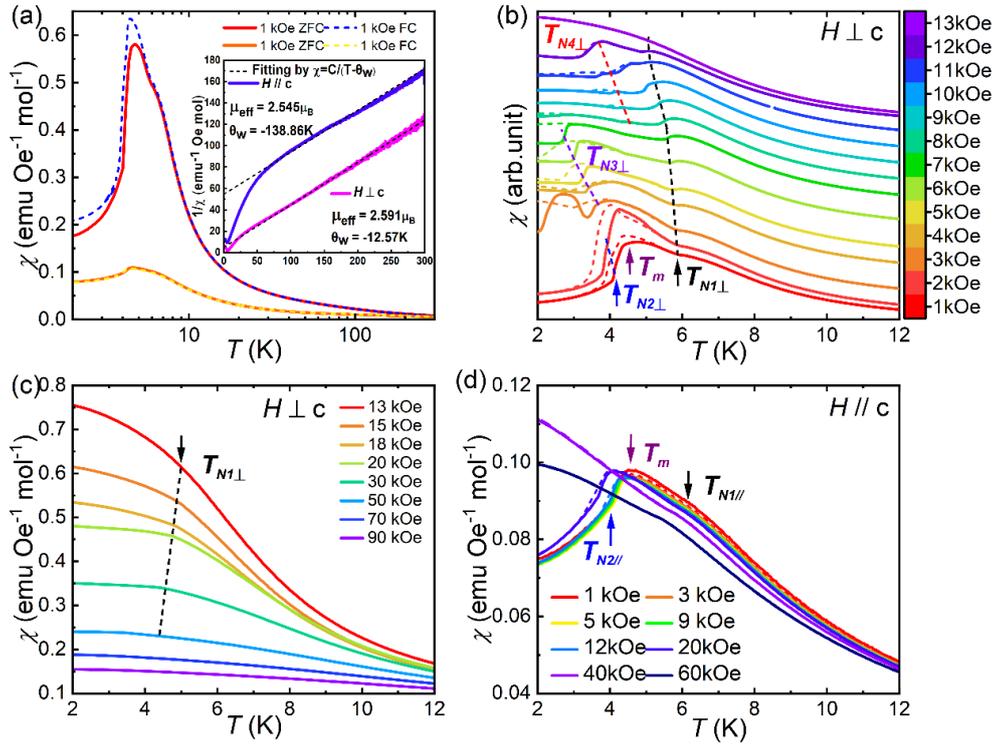

Figure 2. Magnetic susceptibility *(χ)* of $Ce_3ScBi_5$. (a) Temperature-dependent magnetic susceptibility and inverse magnetic susceptibility of $Ce_3ScBi_5$ were collected for $H\perp c$ and $H // c$ (*H* = 1 kOe). (b)-(d) Temperature-dependent magnetic susceptibility under various magnetic fields for $H\perp c$ and $H // c$. Zero-Field Cooling (ZFC) and Field Cooling (FC) data are represented by a solid line and a dashed line, respectively.

***Magnetic orders and magnetization plateaus.*** Temperature-dependent magnetic susceptibility $\chi(T)$, measured with $H \perp c$ and $H // c$, are presented in Figure 2(a). Both $\chi(T)$ curves exhibit a broad hump around 5.9K (denoted as $T_{N1}$) and a sharp kink around 4.1K ($T_{N2}$), indicating two antiferromagnetic (AFM) orderings of the $Ce^{3+}$ moment. Similar magnetic ordering behavior related to two propagation vectors, $k_1 = [0, 1/2, 1/8]$ and $k_2 = [0, 0, 1/8]$, has been reported in powder neutron studies of the isomorphic compound $Ce_3TiSb_5$[31]. Additionally, the zero-field cooling (ZFC) and field cooling (FC) curves for $H \perp c$ show a slight divergence below $T_{N1}$, suggesting a small FM component that may be contributed by the ferromagnetic interactions between $Ce^{3+}$ moments in adjacent layers [32]. The ac magnetic susceptibility curve (Figure S2 in the SM[23]) verifies the absence of spin glass freezing in $Ce_3ScBi_5$. Particularly at low temperatures, the ratio of $\chi_{\perp c}$ to $\chi_{//c}$ can reach a maximum value of ~ 6, reflecting a significant anisotropy where the easily magnetized direction of $Ce_3ScBi_5$ lies perpendicular to the crystallographic c-axis. The corresponding $\chi^{-1}(T)$ curves demonstrate linear behavior above 50 K. The data between 50 and 300 K can allow for good Curie-Weiss fits using $\chi = C/(T-\theta_w)$, which yields the effective magnetic moments $\mu_{eff}$= 2.591 $\mu_B$/Ce, the Weiss temperature $\theta_w$ = –12.57 K for $H \perp c$ and $\mu_{eff}$= 2.545 $\mu_B$/Ce, $\theta_w$ = –138.86 K for $H // c$, respectively. These estimated $\mu_{eff}$ values are both in good accordance with the theoretical moment of 2.54 $\mu_B$ for free $Ce^{3+}$. The negative sign of the $\theta_w$ indicates the antiferromagnetic nature of magnetic interactions between the localized *4f* Ce moments. The $\theta_w$ difference may be owing to a large magnetocrystalline anisotropy from the crystal electric fields (CEFs) [33, 34].

To further investigate the magnetic transition of $Ce_3ScBi_5$, $\chi_{\perp c}(T)$ was measured under different magnetic fields, as illustrated in Figure 2(b, c). At zero field cooling of 1 kOe, three characteristics can be obtained; i.e., the magnetization appeared as a hump near 5.9 K ($T_{N1\perp}$), began to decline near 4.6K ($T_m$), and dropped sharply near 4.1K ($T_{N2\perp}$). The $\chi_{\perp c}(T)$ displays a peak maximum at $T_m$ = 4.6 K, while the thermodynamic determination reveals the Néel temperature to be $T_{N2\perp}$=4.1 K, which is evident from $d\chi/dT$ with respect to temperature in Fig. S4(a) in the SM[23]. This observation suggests that although the static magnetic susceptibility begins to decline below 4.6 K, the dynamic magnetic susceptibility continues to increase below 4.1 K due to spin frustration until long-range antiferromagnetic order occurs. The $\chi_{ac}(T)$ curve further confirms this view. This peculiar magnetic behavior has been observed in other frustrated systems such as CePdAl[35] and $Na_2IrO_3$[36].

Moreover, with an increase in the applied field, $T_{N1}$ decreased, and the hump was no longer observed for fields higher than 50 kOe (More details in $d\chi/dT$ curves can be seen in Figure S4(b) in the SM[23]), indicating a typical AFM behavior. There are three magnetic anomalies worth noting below $T_{N1\perp}$: (1) A kink emerges near $T_{N3\perp}$=3.7 K as $H$ increases to 3 kOe, and $\chi_{\perp c}(T)$ no longer strongly drops near the transition but tends to saturation. (2) The kink around $T_{N3\perp}$ shifts to lower temperatures as $H$ increases until it disappears at 9 kOe, while a new kink emerges near $T_{N4\perp}$=4.5K. The existence of this additional anomaly is also reported in the isomorphic compound $Ce_3ZrBi_5$[37]. (3) The kink around $T_{N4\perp}$ moves towards low temperature as $H$ increases and disappears for fields higher than 12 kOe. The changes in transitions are more clearly demonstrated by

$d\chi/dT$ [Figure S4(a, b) in the SM[23]]. $\chi_{//c}(T)$ curves presented in Figure 2(d) exhibit the same features at $H$=1 kOe, with a hump near 5.9 K ($T_{N1//}$), a peak maximum near 4.6K ($T_m$), and a kink near 4.1K ($T_{N2//}$). As $H$ increases, both transitions at $T_{N1//}$ and $T_{N2//}$ shift slightly towards lower temperatures, while the former persists until $H$=60 kOe and the latter disappears at $H$=40 kOe. In general, the transformations of $Ce_3ScBi_5$ do not evolve monotonically with the field, indicating the existence of multiple magnetic phases.

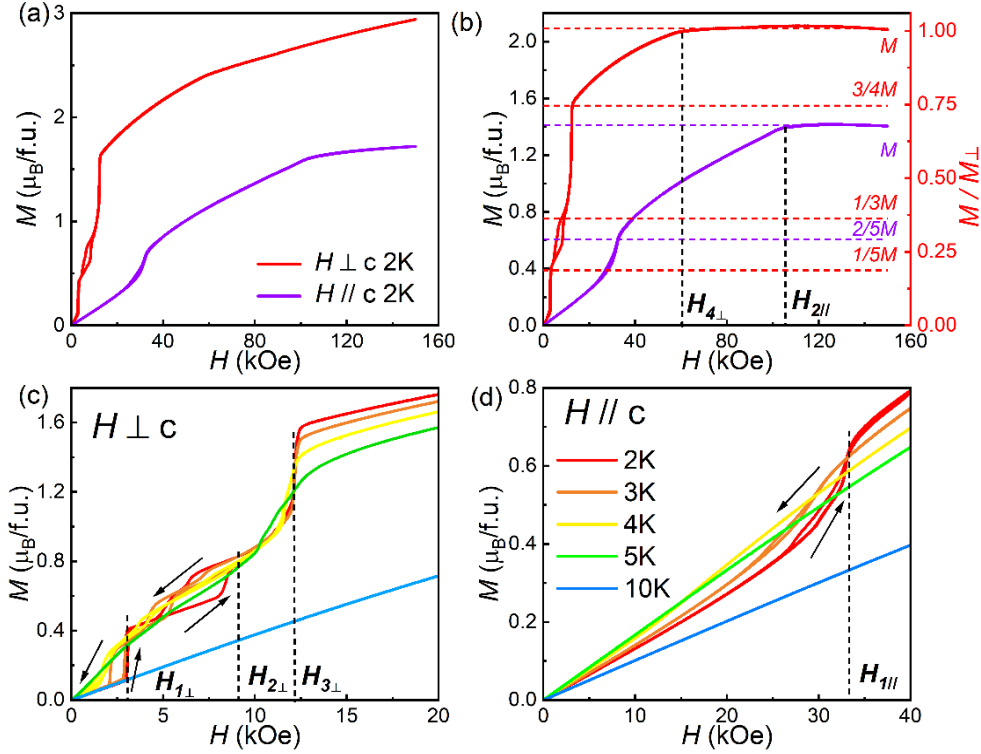

Figure 3. Isothermal magnetizations ($M$) of $Ce_3ScBi_5$. (a) Field-dependent magnetization at $T$ = 2 K for $H\perp c$ and $H // c$. (b) Magnetization curves corrected for the Van Vleck paramagnetism. (c),(d) Isothermal magnetizations under various magnetic fields for $H\perp c$ and $H // c$.

The isothermal magnetization $M(H)$ data at 2 K with $H\perp c$ and $H // c$ are presented in Figure 3(a). At high fields, the data exhibit a linear behavior without saturation up to $H$=150 kOe, indicating the entry of $Ce_3ScBi_5$ into a field-polarized regime attributed to Van Vleck paramagnetism induced by $Ce^{3+}$ cations[38]. After correction for the Van Vleck paramagnetism[39], The $M(H)$ curves in Figure 3(b) reveal the field-induced transitions occurring at $H_{\perp*}$=60 kOe for $H\perp c$ and at $H_{//*}$=105 kOe for $H\perp c$. The values of $M_\perp$ and $M_{//}$ at $H$=150 kOe attain about 0.70 $\mu_B$/Ce and about 0.47 $\mu_B$/Ce, respectively, significantly lower than the free-ion value of $Ce^{3+}$, potentially influenced by the CEF effect. Similar behavior has been observed in other Ce-based intermetallic compounds, such as $CeMg_3$[40] and $CeAgAs_2$[41].

Distinct field-induced transitions can be identified at $H_{1\perp}$= 3 kOe, $H_{2\perp}$= 9 kOe, and $H_{3\perp}$=12 kOe at $T$ = 2 K in Figure 3(c), with the plateau values after the transitions approximately 1/5, 1/3, and 2/3 of $M_\perp$, respectively. The $M_\perp(H)$ curve initially shows a linear increase and then undergoes a jump of about 1/5 $M_\perp$ at critical $H_{1\perp}$, exhibiting narrow hysteresis loops, indicating that the field-induced metamagnetic transition belongs to a first-order phase transition[42]. Moreover, the hysteresis loops nearly disappear around $H_{2\perp}$ and the field-induced hysteretic transition may be associated with the anomaly at $T_{N4\perp}$. The isomorphic compound $La_3MnBi_5$[43] has also been reported to exhibit a similar hysteresis transition, which represents a crossover between two magnetic phases. Then the $M_\perp(H)$ curve undergoes a jump of about 1/3 $M_\perp$ at critical $H_{3\perp}$ and continues to increase after the transition. It is inferred that this metamagnetic transition stabilizes Ce moments into a forced-ferromagnetic state[44]. However, the Ce moments are not perfectly aligned at $H_{3\perp}$ and are further inclined towards the field direction with $H$ increasing. Furthermore, with increasing temperature, the metamagnetic transitions at $H_{1\perp}$ and $H_{3\perp}$ shift towards lower fields [see $dM/dH$ curves in Figure S5(a) in the SM[23]]. Around $H_{1//}$ = 33 kOe, a field-induced transition can be observed for $H // c$ at 2 K from the isothermal magnetizations depicted in Figure 3(d), resembling the hysteresis transition around $H_{2\perp}$. The magnitude of the plateau after the transition near $H_{1//}$ reaches approximately 2/5 of $M_{//}$ and moves to lower fields with $H$ increasing. One possible reason for the emergence of the plateaus is due to the competing AFM and ferromagnetic exchange interactions. Fractional magnetization plateaus serve as clear evidence of the complex spin structures generated by magnetic frustration [45-47].

Through meticulous magnetic measurements, we summarize the magnetic structural characteristics of $Ce_3ScBi_5$. Noncollinearly aligned spins within the *ab*-plane exhibit strong coupling, leading to high magnetic moments along that direction. Under the influence of an external magnetic field, these spins can undergo in-plane rotation with relative ease. Conversely, spin motion along the c-axis is more constrained, making alignment more challenging. It is important to acknowledge that while this model offers substantial insight into the observed magnetism, neutron scattering experiments are imperative for elucidating the precise magnetic arrangement of $Ce_3ScBi_5$.

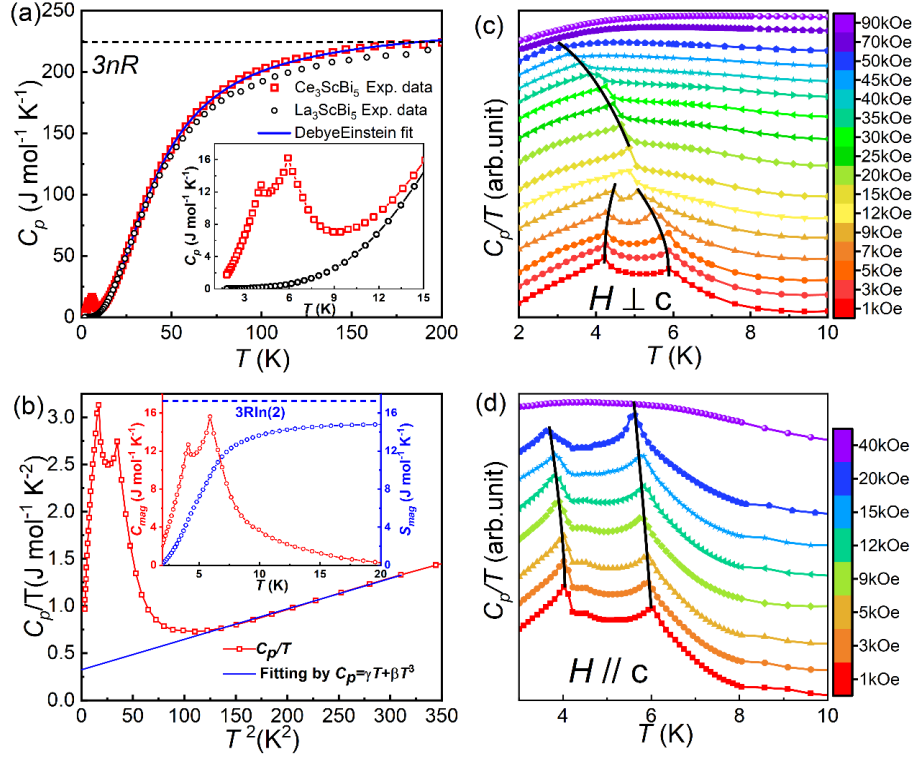

Figure 4. Heat capacity ($C_p$) of $Ce_3ScBi_5$. (a) Temperature-dependent specific heat of $Ce_3ScBi_5$ and $La_3ScBi_5$. The blue curve is the Debye-Einstein fitting. (b) The low-temperature $C_p/T$ versus $T^2$ curve. The inset shows the specific heat and entropy of the magnetic part. The $C_{mag}$ is obtained by subtracting off $C_p(T)$ of $La_3ScBi_5$ from that of $Ce_3ScBi_5$. (c),(d) The low-temperature $C_p/T$ versus $T$ curves with varied magnetic fields for $H\perp c$ and $H // c$.

*Heat capacity.* Temperature-dependent specific heat $C_p(T)$ of $Ce_3ScBi_5$ and its nonmagnetic analog $La_3ScBi_5$ at zero fields are presented in Figure 4(a). The $C_p(T)$ of $Ce_3ScBi_5$ reaches 225.92 J mol$^{-1}$ K$^{-1}$ at 300K, which closely approaches the Dulong-Petit limit [48]. Two consecutive peaks are observed at approximately 4.1 and 5.9 K, confirming the bulk magnetic ordering corresponding to $T_{N1}$ and $T_{N2}$, respectively. Note that an extended tail up to 9 K signifies strong critical spin fluctuation. To further explore the spin state of $Ce_3ScBi_5$, we utilize the Debye-Einstein model[49, 50] to fit the data between 20K and 300K in the $C_p(T)$ curve using the analytical formula:

$$C_p = \gamma T + 9nR\alpha \left(\frac{T}{T_D}\right)^3 \int_0^{\frac{T_D}{T}} \frac{x^4 e^x dx}{(e^x-1)^2} + 3nR(1-\alpha)\left(\frac{T_E}{T}\right)^2 \frac{e^{\frac{T_E}{T}}}{\left(e^{\frac{T_E}{T}}-1\right)^2} \quad (1)$$

in which $\alpha$ represents a weight factor of Debye model and Einstein model. The fit yields $T_D$ =166.99 K, $T_E$=38.72 K and $\gamma^{D-E}$=154.53 mJ mol$_{Ce}^{-1}$K$^{-2}$. The Sommerfeld coefficient $\gamma^{D-E}$ indicates a significant effective electron mass and implies strong electron-correlation effects. Alternatively, the low-temperature data of $Ce_3ScBi_5$ between 10 K and 17 K were excellently fitted to the formula $C_P/T= \gamma+ \beta T^2$. The obtained $\gamma^D$ = 108.26 mJ mol$_{Ce}^{-1}$K$^{-2}$ is in reasonable agreement with the above $\gamma^{D-E}$ value, while the moderately enhanced $\beta$ = 3.23×10$^{-3}$ J mol$^{-1}$ K$^{-4}$ reflects the significant spin wave interaction in $Ce_3ScBi_5$.

The magnetic contribution $C_{mag}$ of Ce and its corresponding magnetic entropy $S_{mag}$ are illustrated in Figure 4(b). The $S_{mag}$ exhibits distinct jumps at both transitions and flattens out to 14.8 J mol$^{-1}$K$^{-1}$ at 20 K, slightly below the theoretical limit for the Kramers doublet ground state of the Ce 4$f$ electrons due to the formation of short-range correlations. Noteworthy, $C_{mag}$ released at $T_{N2} \approx 5.9$ K is only 0.54 × 3Rln2, significantly diminished by spin frustration and Kondo effect [41, 46, 51]. This conclusion is motivated by the magnetic frustration that often arises from competing exchange interactions. Within the mean-field approximation, the reduced entropy at $T_N$ can be used to deduce $T_K \approx 12$ K as a measure of a weak Kondo effect in Ce$_3$ScBi$_5$[52, 53].

Due to the contribution of spin fluctuation, Ce$_3$ScBi$_5$ exhibits a rich phase structure under an external field. The intricate interplay between CEF splitting, Kondo scattering, and RKKY interactions, each operating on closely related energy scales, leads to diverse physical responses under magnetic field application. The temperature $C_p/T$ versus $T$ curves for $H \perp c$ and $H // c$ are presented in Figure 4(c, d). For $H \perp c$, the peak at $T_{N1}$ is gradually suppressed to low temperatures and evolves into a broad crossover, while the peak at $T_{N2}$ initially remains almost invariant before shifting to high temperatures. This anomaly may be ascribed to the influence of a canted antiferromagnetism [54] and the induced metamagnetic transition at critical $H_{1\perp}$ further enhances the FM contribution. Beyond 12 kOe, the peak at $T_{N2}$ converges on the peak at $T_{N1}$, corresponding to a metamagnetic transition at critical $H_{3\perp}$, and then the peak at $T_{N1}$ moves towards lower temperatures with increasing field until it vanishes for fields higher than 50 kOe. It is evident that for $H // c$, the peaks shift towards lower temperatures with increasing fields consistent with the behavior expected for AFM orders.

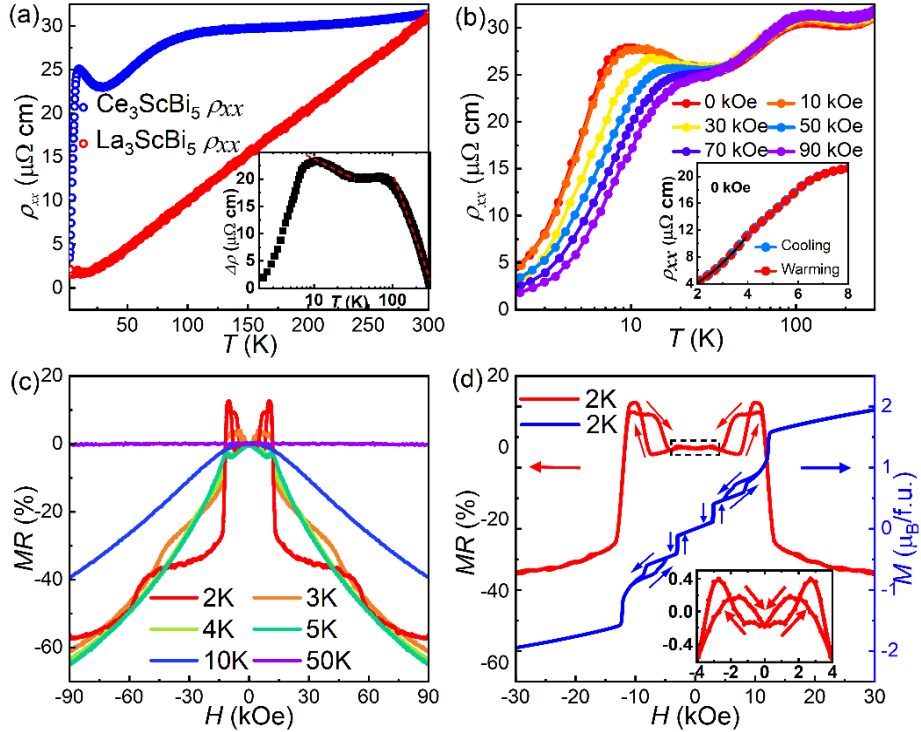

Figure 5. Electronic resistivity ($\rho$) of Ce$_3$ScBi$_5$. (a)Temperature-dependent of $\rho_{xx}(T)$ of Ce$_3$ScBi$_5$ and its nonmagnetic analog La$_3$ScBi$_5$ ($I$ // [001]). The inset is resistivity induced by the magnetism of Ce ions ($\Delta\rho=\rho_{Ce}-\rho_{La}$). (b) Temperature dependence of resistivity for $H\perp$c. The inset shows no thermal hysteresis from the transitions. (c) Isothermal magnetoresistance for $H\perp$c at various temperatures. (d) Field-dependent magnetoresistance (red) and magnetization (blue) curves at 2 K for $H\perp$c. The inset shows details inside the dotted box.

***Charge Transport.*** The $\rho_{xx}(T)$ of Ce$_3$ScBi$_5$ and La$_3$ScBi$_5$ for $H\perp$c are presented in Figure 5(a). The high residual resistivity ratio ($RRR = \rho_{300\,K}/\rho_{2\,K} \approx 9.4$) indicates the exceptional quality of single crystals. The $\rho_{xx}(T)$ displays a broad shoulder at 100K, followed by a plateau with a weak maximum at 10 K, then rapidly decreases below 10 K. The broad shoulder can be possibly ascribed to the cooperation of Kondo scattering and the CEF splitting[34, 40, 55]. The inset of Figure 5(a) shows the magnetic contribution to resistivity, $\Delta\rho=\rho_{Ce}-\rho_{La}$, on a logarithmic scale, with two humps at around 10 K and 100 K, and an almost linear dependence on $lnT$ in the subsequent temperature regions. This behavior aligns with Kondo scattering in a strong crystal field as described by Cornut and Coqblin models [56]. The hump observed at 10 K corresponds closely to $T_K$ obtained previously and is primarily influenced by Kondo interactions.

The electrical transport of Ce$_3$ScBi$_5$ reveals Kondo-lattice characteristics. Figure 5(b) illustrates $\rho_{xx}(T)$ curves for $H\perp$c under various magnetic fields, showcasing the absence of hysteresis in the transitions, thereby suggesting that transitions are governed by magnetic rather than structural factors. The Fermi-liquid behavior following $\rho(T) = \rho_0 + AT^n$ can be confirmed below $T_{N1}$, where $\rho_0$ = 1.788 μΩ cm, A = 0.5896 μΩ cm K$^{-2}$ and n = 2.009. At zero fields, it is difficult to discern the signature of magnetic ordering in $\rho_{xx}(T)$, while two broad peaks were observed at transitions of the $d\rho/dT$ curve [Figure S6(a) in the SM[23]]. It is worth noting that $\rho_{0xx}$ decreases as the field increases, presumably due to the suppression of spin fluctuations by high fields [41]. Furthermore, the metallic behavior exhibited in $\rho_{zz}(T)$ with $I$ // [100] [Figure S7(a, b) in the SM[23]] highlights a substantial resistivity anisotropy at low temperatures, with $\rho_{zz}/\rho_{xx} \approx 16$. Upon applying a field up to 40 kOe, the $\rho_{zz}(T)$ profile becomes smooth, and the transitions move towards low temperatures, as also evidenced in $\chi(T)$ and $C_p(T)$ curves.

***Anisotropy Magnetoresistance and Hall effect Anomalies.*** Isothermal *MR* for $H\perp$c at various temperatures, as depicted in Figure 5(c), illustrates a negative colossal *MR* of up to 60% at 90 kOe in low-temperature regimes. Analogous negative colossal *MR* behavior has been observed in magnetic systems such as MnSbTe[50], CrSCr[57] and EuMnSb$_2$[58]. It is suggested that the conduction by minority spin carriers could account for colossal *MR*[50, 57]. The *MR* exhibits a tendency towards diminution as temperature rises, evolving into a quasilinear trend above 10K. This linear *MR* has been reported in LaAgBi$_2$[59], possibly due to the potential Dirac dispersion of Bi itinerant electrons. It is noted that a sharp drop near 60 kOe at 2 K further confirms the existence of a field-induced transition at $H_{\perp*}$ in $M(H)$ curves. Upon closer analysis of *MR* and *MH* curves at 2 K [Figure 5(d)], it is observed that the *MR* profiles develop two distinct hysteresis loops, diminishing in magnitude as temperature increases, in line with the formation of magnetic hysteresis. This hysteresis loop indicates a coexistence state of two phases, where the antiferromagnetic component contributes to positive *MR*[60]. Subsequently, the large negative *MR* at 12 kOe is primarily ascribed to spin disorder quenching upon applying an external magnetic field[61, 62] and aligns with the metamagnetic transition at critical $H_{3\perp}$. The exhibited characteristics in *MR* closely correlate with the *M(H)* behavior, leading to the inference that the electrical properties of $Ce_3ScBi_5$ are intricately interconnected with its magnetic order.

Most importantly, a significant anisotropy magnetoresistance of $Ce_3ScBi_5$ was observed at low temperatures. The isothermal *MR* of *H // c* drops sharply at around 31 kOe at 2 K, exhibiting a slight hysteresis attributed to the field-induced transition [Figure S7(c,d) in the SM[23]]. When plotting *MR* as a function of angle [Figure 6(a)], the metamagnetic features are most pronounced at $\theta = 0°$ and highly sensitive to the field angle, indicating a strong anisotropic *MR* in this system. It is noteworthy that the critical field for the sharp drop in *MR* monotonically increases with the angle up to 90°. For improved visualization, the angle-dependent *MR* is represented in polar coordinates in Figure 6(b), from which we can observe maximum at angles of 60°, 120°, 240°, and 300°. This is possibly due to a change in Fermi surface geometry along this direction. These data establish that the transitions are most sensitive to the contribution of the magnetic field.

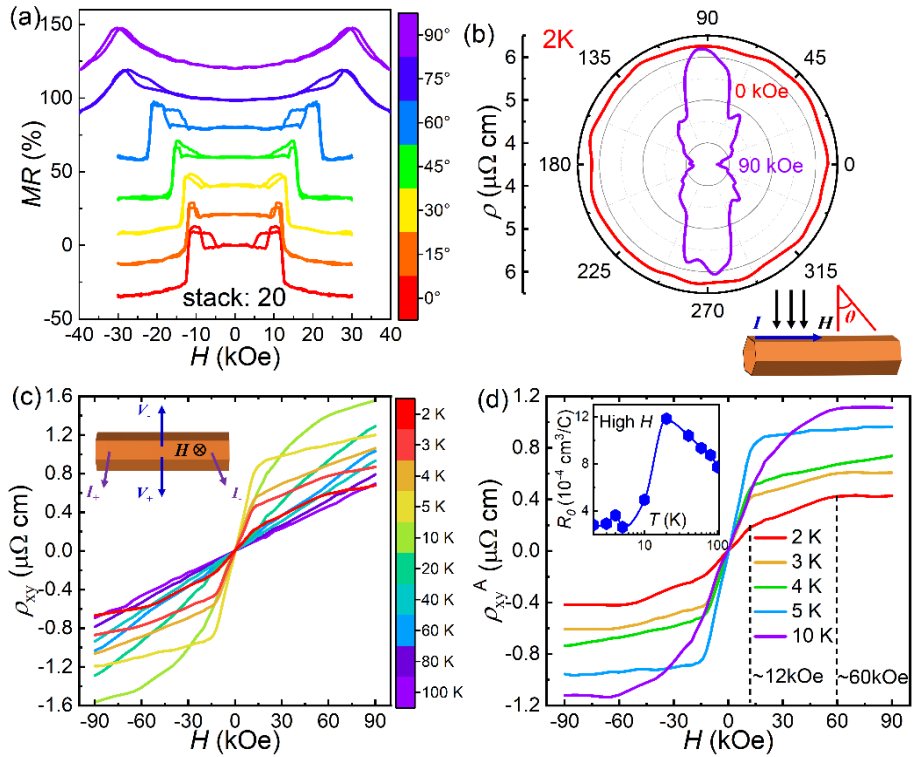

Figure 6. Angle-dependent magnetoresistance and Hall effect of $Ce_3ScBi_5$. (a) *MR* as a function of $H \perp c$ at different tilt angles θ (*I //* [001]). (b) Directional dependence of the resistivity at 2 K. (c) Field-dependent Hall resistivity $\rho_{xy}$ measured at constant different temperatures. (d) The anomalous Hall resistivity $\rho^A_{xy}$ extracted from $\rho_{xy}$. The inset shows the temperature dependence of $R_0(T)$.

To gain deeper insights into the intriguing behavior of transport properties, Figure 6(c) presents the Hall resistivity $\rho_{xy}$, which exhibits a linear correlation above 20 K and decreases with increasing temperature. At 100 K, the observed $\rho_{xy}$ is predominantly influenced by the ordinary Hall effect, implying an estimated carrier density of approximately $8.1\times10^{17}$ cm$^{-3}$. However, below 10 K, $\rho_{xy}$ displays non-linear behavior, indicating the presence of Hall effect anomalies. The subtraction of the linear component from the Hall resistivity yields the anomalous Hall resistivity $\rho^A_{xy}$, as depicted in Figure 6(d). $R_0$ consistently remains positive, indicating the dominant role of holes in transport. At low temperatures, $\rho^A_{xy}$ exhibits a sharp increase and changes slope at 12 kOe before plateauing at 60 kOe, which is consistent with the behavior observed by *M(H)* and *MR* curves. The Hall data shows obvious sensitivity to field-induced transition, and one possible explanation is that field-induced magneto-polarons contribute to anomalies[63, 64]. Further analysis and discussion in the future are necessary to explore the origin of Hall effect anomalies.

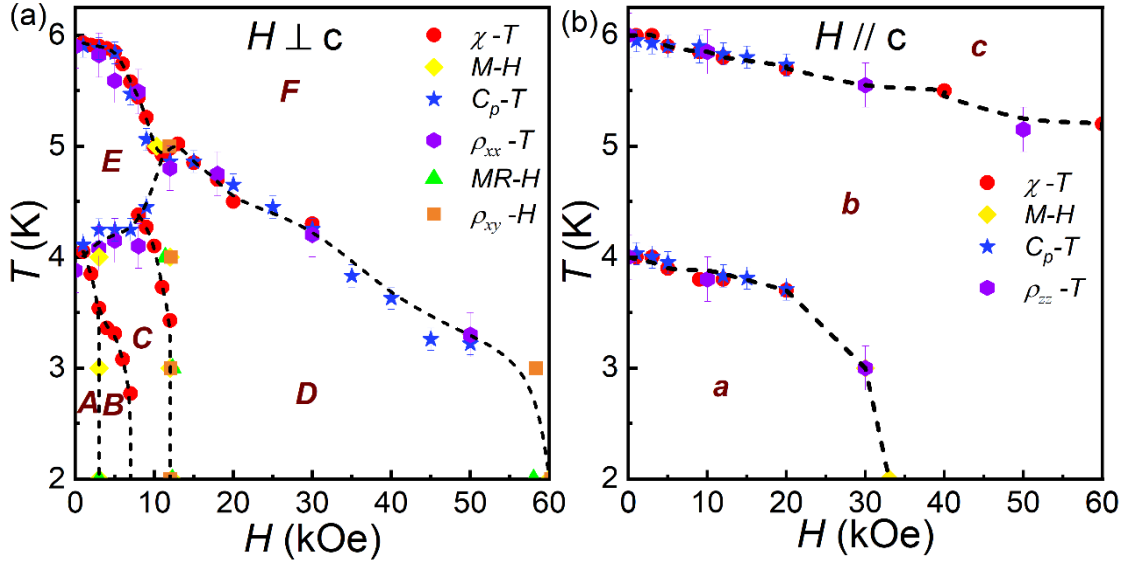

Figure 7. Temperature-magnetic field phase diagram of Ce$_3$ScBi$_5$ with the applied field both perpendicular (a) and parallel (b) to the c axis.

***Temperature-magnetic field phase diagram.*** Due to the sensitivity of spin frustration to the external field, a multi-step spin reorientation transition is observed, particularly in the $H\perp$c phase diagram depicted in Figure 7(a). Based on the aforementioned analysis, we can infer that: the A phase represents the canted-AFM II state, followed by the metamagnetic transition that occurs at 3kOe, resulting in the field-induced B and C phases corresponding to AFM II-1 and AFM II-2 states, respectively. Likewise, the E phase represents the AFM I state, followed by another metamagnetic transition that occurs at 12Oe (e.g., T = 5 K), when the AFM II state disappears, and AFM I is induced into the D phase (AFMI-1). It is worth noting that near 6 K at the boundary of the E phase, there is a smooth decrease before reaching the critical field which signifies frustration-induced quantum fluctuation. The boundaries between the B and C phases cannot be determined through specific heat due to minimal entropy involvement. The discrepancy between $T_N$ values obtained from $\chi(T)$ and $C_p(T)$ curves serves as evidence for competition between antiferromagnetic states and short-range orders[35, 47, 65]. In contrast, the $H // $ c phase diagram shown in Figure 7(b) provides clearer information where it becomes evident that the *a* phase corresponds to the AFM II state while the *b* phase corresponds to the AFM I state.

### IV. CONCLUSIONS

In summary, we have successfully synthesized a high-quality single crystal of $Ce_3ScBi_5$, a recently identified compound within the $Ln_3MX_5$ family. The ground-state magnetism and electrical transport affected by the cooperative competition of geometric frustration, Kondo physics, and RKKY interactions are systematically investigated. Magnetization results reveal that $Ce_3ScBi_5$ undergoes two antiferromagnetic transitions at temperatures $T_{N1}$=5.9 K and $T_{N2}$=4.1 K. Moreover, for $H\perp$c orientation, it displays two metamagnetic transitions, showcasing the fractional magnetization platform. The complex spin structure generated by magnetic frustration under the action of an external field leads to a multistep spin reorientation transition.

Heat capacity and electrical resistivity measurements further support this view and indicate that $Ce_3ScBi_5$ is a Kondo-lattice system with a magnetic ground state. Of note, $Ce_3ScBi_5$ possesses a particularly intriguing transport behavior, attributed to its pronounced anisotropic magnetoresistance, which holds significant promise for material applications. It is found that there exists a strong correlation between magnetic order and electrical transport. Additionally, a significant anomalous Hall effect was detected at low temperatures. The phase diagrams determined from our measurements indicate that $Ce_3ScBi_5$ is strongly anisotropic and exhibits a rich variety of magnetic states arising from competitive interactions. Quantum fluctuations, Kondo coherence, and antiferromagnetic ordering at close energy scales produce complex responses of physical properties to the application of magnetic field. Overall, our findings establish $Ce_3ScBi_5$ as a promising platform for studying geometric frustration in Kondo lattices while also opening up more avenues for experimental and theoretical investigations such as antiferromagnetic quantum critical behaviors.

**ACKNOWLEDGMENTS**

This work was supported by the National Key Research and Development Program of China (Grant No. 2021YFA1400401, No. 2023YFA1607400), the National Natural Science Foundation of China (Grants No. 52271238, No.12274436, No. U22A6005, No. U2032204), the Strategic Priority Research Program of the Chinese Academy of Sciences (Grant No. XDB33010000), the Center for Materials Genome, and the SynergeticExtreme Condition User Facility (SECUF)

## References:

[1] L. Balents, Spin liquids in frustrated magnets, *Nature*. **464**, 199 (2010).

[2] A. P. Ramirez, A. Hayashi, R. J. Cava, R. Siddharthan, B. S. Shastry, Zero-point entropy in 'spin ice', *Nature*. **399**, 333 (1999).

[3] J. Schulenburg, A. Honecker, J. Schnack, J. Richter, H. J. Schmidt, Macroscopic Magnetization

Jumps due to Independent Magnons in Frustrated Quantum Spin Lattices, *Phys. Rev. Lett.* **88**, 167207 (2002).

[4] Q. Si, Quantum criticality and global phase diagram of magnetic heavy fermions, *Phys. Status Solidi B*. **247**, 476 (2010).

[5] R. Movshovich, M. Jaime, J. D. Thompson, C. Petrovic, Z. Fisk, P. G. Pagliuso, J. L. Sarrao, Unconventional Superconductivity in CeIrIn$_5$ and CeCoIn$_5$: Specific Heat and Thermal Conductivity Studies, *Phys. Rev. Lett.* **86**, 5152 (2001).

[6] J. Custers, P. Gegenwart, H. Wilhelm, K. Neumaier, Y. Tokiwa, O. Trovarelli, C. Geibel, F. Steglich, C. Pépin, P. Coleman, The break-up of heavy electrons at a quantum critical point, *Nature*. **424**, 524 (2003).

[7] L. Ye, S. Fang, M. Kang, J. Kaufmann, Y. Lee, C. John, P. M. Neves, S. Y. F. Zhao, J. Denlinger, C. Jozwiak, *et al.*, Hopping frustration-induced flat band and strange metallicity in a kagome metal, *Nature Physics*. (2024).

[8] W. J. Kim, M. A. Smeaton, C. Jia, B. H. Goodge, B.-G. Cho, K. Lee, M. Osada, D. Jost, A. V. Ievlev, B. Moritz, *et al.*, Geometric frustration of Jahn–Teller order in the infinite-layer lattice, *Nature*. **615**, 237 (2023).

[9] M. Xu, L. H. Kendrick, A. Kale, Y. Gang, G. Ji, R. T. Scalettar, M. Lebrat, M. Greiner, Frustration- and doping-induced magnetism in a Fermi–Hubbard simulator, *Nature*. **620**, 971 (2023).

[10] Y. Tokiwa, M. Garst, P. Gegenwart, S. L. Bud'ko, P. C. Canfield, Quantum Bicriticality in the Heavy-Fermion Metamagnet YbAgGe, *Phys. Rev. Lett.* **111**, 116401 (2013).

[11] Q. Faure, S. Takayoshi, S. Petit, V. Simonet, S. Raymond, L.-P. Regnault, M. Boehm, J. S. White, M. Månsson, C. Rüegg, *et al.*, Topological quantum phase transition in the Ising-like antiferromagnetic spin chain BaCo$_2$V$_2$O$_8$, *Nature Physics*. **14**, 716 (2018).

[12] Z. Dun, M. Daum, R. Baral, H. E. Fischer, H. Cao, Y. Liu, M. B. Stone, J. A. Rodriguez-Rivera, E. S. Choi, Q. Huang, *et al.*, Neutron scattering investigation of proposed Kosterlitz-Thouless transitions in the triangular-lattice Ising antiferromagnet TmMgGaO$_4$, *Phys. Rev. B*. **103**, 064424 (2021).

[13] G. Bollore, M. J. Ferguson, R. W. Hushagen, A. Mar, New ternary rare-earth transition-metal antimonides RE$_3$MSb$_5$ (RE=La, Ce, Pr, Nd, Sm; M=Ti, Zr, Hf, Nb), *Chem. Mater.* **7**, 2229 (1995).

[14] S. H. D. Moore, L. Deakin, M. J. Ferguson, A. Mar, Physical properties and bonding in RE$_3$TiSb$_5$ (RE = La, Ce, Pr, Nd, Sm), *Chem. Mater.* **14**, 4867 (2002).

[15] O. Y. Zelinska, A. Mar, Ternary rare-earth manganese bismuthides:Structures and physical properties of RE$_3$MnBi$_5$ (RE = La-Nd) and Sm$_2$Mn$_3$Bi$_6$, *Inorg. Chem.* **47**, 297 (2008).

[16] J. F. Khoury, X. Song, L. M. Schoop, Ln$_3$MBi$_5$ (Ln=Pr, Nd, Sm; M=Zr, Hf): Intermetallics with Hypervalent Bismuth Chains, *Z. Anorg. Allg. Chem.* **648**, e202200123 (2022).

[17] M. Matin, R. Kulkarni, A. Thamizhavel, S. K. Dhar, A. Provino, P. Manfrinetti, Probing the magnetic ground state of single crystalline Ce$_3$TiSb$_5$, *J. Phys.: Condens. Matter*. **29**, 145601 (2017).

[18] G. Motoyama, M. Sezaki, J. Gouchi, K. Miyoshi, S. Nishigori, T. Mutou, K. Fujiwara, Y. Uwatoko, Magnetic properties of new antiferromagnetic heavy-fermion compounds, Ce$_3$TiBi$_5$ and CeTi$_3$Bi$_4$, *Physica B*. **536**, 142 (2018).

[19] G. Motoyama, M. Shinozaki, S. Nishigori, A. Yamaguchi, N. Aso, T. Mutou, M. Manago, K. Fujiwara, A. Sumiyama, Y. Uwatoko, Transport, Thermal, and Magnetic Properties of Heavy Fermion Compound Ce$_3$TiBi$_5$, *JPS Conf. Proc.* **38**, 011084 (2023).

[20] L. Duan, X. C. Wang, J. Zhang, Z. Hu, J. F. Zhao, Y. G. Feng, H. L. Zhang, H. J. Lin, C. T. Chen,

W. Wu, *et al.*, Synthesis, structure, and magnetism in the ferromagnet La$_3$MnAs$_5$ : Well-separated spin chains coupled via itinerant electrons, *Phys. Rev. B*. **106**, 184405 (2022).

[21] L. Duan, X. Wang, F. Zhan, J. Zhang, Z. Hu, J. Zhao, W. Li, L. Cao, Z. Deng, R. Yu, *et al.*, High-pressure synthesis, crystal structure and physical properties of a new Cr-based arsenide La$_3$CrAs$_5$, *Sci. China. Mater.* **63**, 1750 (2020).

[22] J. F. Khoury, B. Han, M. Jovanovic, R. Singha, X. Song, R. Queiroz, N.-P. Ong, L. M. Schoop, A Class of Magnetic Topological Material Candidates with Hypervalent Bi Chains, *J. Am. Chem. Soc.* **144**, 9785 (2022).

[23] See Supplemental Material at [] for additional experimental information; single crystal structure refinement and characterization details; supplementary magnetic data including magnetization, magnetic susceptibility and their derivatives; supplementary the derivative of resistivity and isothermal magnetoresistance.

[24] Z. M. Sun, J. G. Mao, Synthesis and crystal structure of EuBi$_2$, *J. Solid State Chem.* **177**, 3752 (2004).

[25] R. Nesper, The Zintl-Klemm Concept–A Historical Survey, *Z. Anorg. Allg. Chem.* **640**, 2639 (2014).

[26] G. A. Papoian, R. Hoffmann, Hypervalent Bonding in One, Two, and Three Dimensions: Extending the Zintl-Klemm Concept to Nonclassical Electron-Rich Networks, *Angew. Chem. Int. Ed.* **39**, 2408 (2000).

[27] T. Murakami, T. Yamamoto, F. Takeiri, K. Nakano, H. Kageyama, Hypervalent Bismuthides La$_3$MBi$_5$ (M = Ti, Zr, Hf) and Related Antimonides: Absence of Superconductivity, *lnorg. Chem.* **56**, 5041 (2017).

[28] A. O. Pecharsky, K. A. Gschneidner, Crystal structure and magnetic properties of the new ternary compound Ce$_3$MnBi$_5$, *J. Alloys Compd.* **287**, 67 (1999).

[29] M. D. Núñez-Regueiro, C. Lacroix, B. Canals, Magnetic ordering in the frustrated Kondo lattice compound CePdAl, *Physica C: Superconductivity*. **282-287**, 1885 (1997).

[30] J. Zhang, H. Zhao, M. Lv, S. Hu, Y. Isikawa, Y. F. Yang, Q. Si, F. Steglich, P. Sun, Kondo destruction in a quantum paramagnet with magnetic frustration, *Phys. Rev. B*. **97**, 235117 (2018).

[31] C. Ritter, A. K. Pathak, R. Filippone, A. Provino, S. K. Dhar, P. Manfrinetti, Magnetic ground states of Ce$_3$TiSb$_5$, Pr$_3$TiSb$_5$ and Nd$_3$TiSb$_5$ determined by neutron powder diffraction and magnetic measurements, *J. Phys.: Condens. Matter*. **33**, 245801 (2021).

[32] X. Han, Y. Li, M. Yang, S. Miao, D. Yan, Y. Shi, Complex magnetic and transport properties in Pr$_3$MgBi$_5$: A material with distorted kagome lattice, *Physical Review Materials*. **7**, 124406 (2023).

[33] M. O. Ajeesh, T. Shang, W. B. Jiang, W. Xie, R. D. Dos Reis, M. Smidman, C. Geibel, H. Q. Yuan, M. Nicklas, Ising-type Magnetic Anisotropy in CePd$_2$As$_2$, *Sci.Rep.* **7**, 7338 (2017).

[34] F. Han, X. Wan, D. Phelan, C. C. Stoumpos, M. Sturza, C. D. Malliakas, Q. a. Li, T. H. Han, Q. Zhao, D. Y. Chung, *et al.*, Antiferromagnetic Kondo lattice in the layered compound CePd$_{1-x}$Bi$_2$ and comparison to the superconductor LaPd$_{1-x}$Bi$_2$, *Phys. Rev. B*. **92**, 045112 (2015).

[35] A. Oyamada, S. Maegawa, M. Nishiyama, H. Kitazawa, Y. Isikawa, Ordering mechanism and spin fluctuations in a geometrically frustrated heavy-fermion antiferromagnet on the Kagome-like lattice CePdAl: A $^{27}$Al NMR study, *Phys. Rev. B*. **77**, 064432 (2008).

[36] Y. Singh, S. Manni, J. Reuther, T. Berlijn, R. Thomale, W. Ku, S. Trebst, P. Gegenwart, Relevance of the Heisenberg-Kitaev Model for the Honeycomb Lattice Iridates A$_2$IrO$_3$, *Phys. Rev. Lett.* **108**, 127203 (2012).

[37] G. Motoyama, M. Shinozaki, M. Tsubouchi, M. Kuninaka, S. Nishigori, K. Miyoshi, K. Fujiwara,

J. Gouchi, Y. Uwatoko, Magnetic Properties of New Antiferromagnetic Compound of $Ce_3ZrBi_5$, *JPS Conf. Proc.* **30**, 011180 (2020).

[38] M. M. Bordelon, C. Girod, F. Ronning, K. Rubi, N. Harrison, J. D. Thompson, C. dela Cruz, S. M. Thomas, E. D. Bauer, P. F. S. Rosa, Interwoven atypical quantum states in $CeLiBi_2$, *Phys. Rev. B*. **106**, 214433 (2022).

[39] T. Susuki, N. Kurita, T. Tanaka, H. Nojiri, A. Matsuo, K. Kindo, H. Tanaka, Magnetization Process and Collective Excitations in the S=1/2 Triangular-Lattice Heisenberg Antiferromagnet $Ba_3CoSb_2O_9$, *Phys. Rev. Lett.* **110**, 267201 (2013).

[40] P. K. Das, N. Kumar, R. Kulkarni, A. Thamizhavel, Magnetic properties of the heavy-fermion antiferromagnet $CeMg_3$, *Phys. Rev. B*. **83**, 134416 (2011).

[41] R. Mondal, R. Bapat, S. K. Dhar, A. Thamizhavel, Magnetocrystalline anisotropy in the Kondo-lattice compound $CeAgAs_2$, *Phys. Rev. B*. **98**, 115160 (2018).

[42] K. Feng, I. A. Leahy, O. Oladehin, K. Wei, M. Lee, R. Baumbach, Magnetic ordering in $GdAuAl_4Ge_2$ and $TbAuAl_4Ge_2$: Layered compounds with triangular lanthanide nets, *J. Magn. Magn. Mater.* **564**, 170006 (2022).

[43] C. Zhang, Y. Wang, J. Zheng, L. Du, Y. Li, X. Han, E. Liu, Q. Wu, Y. Shi, Crystal growth, transport, and magnetic properties of quasi-one-dimensional $La_3MnBi_5$, *Physical Review Materials*. **8**, 034402 (2024).

[44] Y. Takahashi, T. Urata, H. Ikuta, Metamagnetic transitions associated with two antiferromagnetic phases of $PrMn_{1-x}Sb_2$ and its magnetic phase diagram, *Phys. Rev. B*. **104**, 054408 (2021).

[45] K. Zhao, H. Deng, H. Chen, K. A. Ross, V. Petricek, G. Gunther, M. Russina, V. Hutanu, P. Gegenwart, Realization of the kagome spin ice state in a frustrated intermetallic compound, *Science*. **367**, 1218 (2020).

[46] Z. Zhuang, M. Lyu, T. Zhang, X. Zhang, Z. Wang, H. Zhao, J. Xiang, Y. Isikawa, S. Zhang, P. Sun, Magnetism in frustrated Kondo and non-Kondo intermetallics: $CeInCu_2$ versus $NdInCu_2$, *Phys. Rev. B*. **107**, 195154 (2023).

[47] H. Zhao, J. Zhang, M. Lyu, S. Bachus, Y. Tokiwa, P. Gegenwart, S. Zhang, J. Cheng, Y.-f. Yang, G. Chen*, et al.*, Quantum-critical phase from frustrated magnetism in a strongly correlated metal, *Nature Physics*. **15**, 1261 (2019).

[48] C. Kittel, *Introduction to Solid State Physics*,(Wiley, New York, 1996)

[49] C. L. Yang, X. Wang, X. Zhang, D. S. Wu, M. Liu, P. Zheng, J. Y. Yao, Z. Z. Li, Y. F. Yang, Y. G. Shi*, et al.*, Kondo effect in the quasiskutterudite $Yb_3Os_4Ge_{13}$, *Phys. Rev. B*. **91**, 075120 (2015).

[50] D. Y. Yan, M. Yang, P. B. Song, Y. T. Song, C. X. Wang, C. J. Yi, Y. G. Shi, Site mixing induced ferrimagnetism and anomalous transport properties of the Weyl semimetal candidate $MnSb_2Te_4$, *Phys. Rev. B*. **103**, 224412 (2021).

[51] M. Lyu, H. Zhao, J. Zhang, Z. Wang, S. Zhang, P. Sun, $CeAu_2In_4$: A candidate of quasi-one-dimensional antiferromagnetic Kondo lattice, *Chinese Physics B*. **30**, 087101 (2021).

[52] M. O. Ajeesh, S. K. Kushwaha, S. M. Thomas, J. D. Thompson, M. K. Chan, N. Harrison, J. M. Tomczak, P. F. S. Rosa, Localized f-electron magnetism in the semimetal $Ce_3Bi_4Au_3$, *Phys. Rev. B*. **108**, 245125 (2023).

[53] J. A. Blanco, M. De Podesta, J. I. Espeso, J. C. Gómez Sal, C. Lester, K. A. Mcewen, N. Patrikios, J. Rodríguez Fernández, Specific heat of $CeNi_xPt_{1-x}$ pseudobinary compounds and related dilute alloys, *Phys. Rev. B*. **49**, 15126 (1994).

[54] Y. Li, M. Yang, D. Yan, S. Miao, H. Yang, H. L. Feng, Y. Shi, Crystal growth and magnetic properties

of LaMn$_{0.91}$Sb$_2$ and NdMn$_{0.88}$Sb$_2$, *Phys. Rev. B*. **105**, 224429 (2022).

[55] L. Wang, C. Wang, Z. Liu, J. Cheng, S. Miao, Y. Song, Y. Shi, Y. F. Yang, Magnetic phase diagrams of the ferromagnetic Kondo lattice CePd$_2$Al$_8$, *Phys. Rev. B*. **100**, 085122 (2019).

[56] B. Cornut, B. Coqblin, Influence of the Crystalline Field on the Kondo Effect of Alloys and Compounds with Cerium Impurities, *Phys. Rev. B*. **5**, 4541 (1972).

[57] E. J. Telford, A. H. Dismukes, K. Lee, M. Cheng, A. Wieteska, A. K. Bartholomew, Y. S. Chen, X. Xu, A. N. Pasupathy, X. Zhu, *et al.*, Layered Antiferromagnetism Induces Large Negative Magnetoresistance in the van der Waals Semiconductor CrSBr, *Adv. Mater.* **32**, 2003240 (2020).

[58] C. Yi, S. Yang, M. Yang, L. Wang, Y. Matsushita, S. Miao, Y. Jiao, J. Cheng, Y. Li, K. Yamaura, *et al.*, Large negative magnetoresistance of a nearly Dirac material: Layered antimonide EuMnSb$_2$, *Phys. Rev. B*. **96**, 205103 (2017).

[59] K. Wang, D. Graf, C. Petrovic, Quasi-two-dimensional Dirac fermions and quantum magnetoresistance in LaAgBi$_2$, *Phys. Rev. B*. **87**, 235101 (2013).

[60] H. Yamada, S. Takada, Magnetoresistance of Antiferromagnetic Metals Due to s-d Interaction, *J. Phys. Soc. Jpn.* **34**, 51 (1973).

[61] J. Hu, T. F. Rosenbaum, J. B. Betts, Current Jets, Disorder, and Linear Magnetoresistance in the Silver Chalcogenides, *Phys. Rev. Lett.* **95**, 186603 (2005).

[62] A. Pandey, C. Mazumdar, R. Ranganathan, D. C. Johnston, Multiple crossovers between positive and negative magnetoresistance versus field due to fragile spin structure in metallic GdPd$_3$, *Sci.Rep.* **7**, 42789 (2017).

[63] M. Pohlit, S. Rossler, Y. Ohno, H. Ohno, S. von Molnar, Z. Fisk, J. Muller, S. Wirth, Evidence for Ferromagnetic Clusters in the Colossal-Magnetoresistance Material EuB$_6$, *Phys. Rev. Lett.* **120**, 257201 (2018).

[64] P. Rosa, Y. Xu, M. Rahn, J. Souza, S. Kushwaha, L. Veiga, A. Bombardi, S. Thomas, M. Janoschek, E. Bauer, *et al.*, Colossal magnetoresistance in a nonsymmorphic antiferromagnetic insulator, *npj Quantum Materials*. **5**, 52 (2020).

[65] H. Zhao, J. Zhang, S. Hu, Y. Isikawa, J. Luo, F. Steglich, P. Sun, Temperature-field phase diagram of geometrically frustrated CePdAl, *Phys. Rev. B*. **94**, 235131 (2016).

# Supplementary materials

Zhongchen Xu, [1,2] Yaxian Wang[3#], Cuiwei Zhang, [2] Hongxiong Liu, [2] Xin Han, [2] Xianmin Zhang, [1*] Youguo Shi, [2,3,4*] and Quansheng Wu,[3,4*]

[1]Key Laboratory for Anisotropy and Texture of Materials (Ministry of Education), School of Material Science and Engineering, Northeastern University, Shenyang 110819, China

[2] Center of Materials Science and Optoelectronics Engineering, University of Chinese Academy of Sciences, Beijing 100190, China

[3]Beijing National Laboratory for Condensed Matter Physics and Institute of Physics, Chinese Academy of Sciences, Beijing 100190, China

[4]Songshan Lake Materials Laboratory, Dongguan, Guangdong 523808, China

To whom correspondence should be addressed.
[*]ygshi@iphy.ac.cn
[*]yaxianw@iphy.ac.cn
[*]zhangxm@atm.neu.edu.cn

Supporting materials include the following:

1. Additional Experimental Information.
2. Table S1 shows crystallographic and structure refinement data of $Ce_3ScBi_5$.
3. Table S2 shows atomic coordinates and equivalent isotropic thermal parameters of $Ce_3ScBi_5$.
4. Table S3 shows anisotropic atomic displacement parameters ($Å^2$) for $Ce_3ScBi_5$.
5. Figure S1 shows the isothermal magnetization of $Ce_3ScBi_5$ measured at 2K in the ab plane.
6. Figure S2 shows the ac magnetic susceptibility of $Ce_3ScBi_5$ measured in a fixed magnetic field of 10 Oe.
7. Figure S3 shows characterization of $Ce_3ScBi_5$.
8. Figure S4 shows the derivative of magnetic susceptibility ($d\chi/dT$) for $Ce_3ScBi_5$.
9. Figure S5 shows the $dM/dH$ curves for $Ce_3ScBi_5$ measured at different temperatures.
10. Figure S6 shows the derivative of resistivity ($d\rho/dT$) for $Ce_3ScBi_5$ under various magnetic fields.
11. Figure S7 shows $\rho_{zz}(T)$ curves and isothermal magnetoresistance of $Ce_3ScBi_5$.

**Additional Experimental Information**

Single crystals of $Ce_3ScBi_5$ were mounted on Kapton loops with Paratone oil for single crystal X-ray diffraction (SCXRD). Diffraction data were conducted on a Bruker D8 Venture diffractometer at 274 (2) K using Mo Kα radiation (λ = 0.71073 Å). The structures were solved via SHELXT using intrinsic phasing and refined with SHELXL utilizing the least-squares method[23]. The diffraction peaks corresponding to the (*h*00) surface of single crystals were obtained via a Bruker D2 phaser XRD detector by using Cu Kα1 radiation (λ = 1.54184 Å).

Scanning electron microscopy (SEM) and energy dispersive X-ray (EDX) analysis were performed using a Hitachi S-4800 scanning electron microscope with an accelerated voltage of 15 kV. The compounds are phase-pure, with no significant deviations from the $Ce_3ScBi_5$ stoichiometry observed and more characteristic details can be seen in Figure S2(b-g).

Magnetic susceptibility and isothermal magnetizations of $Ce_3ScBi_5$ were measured in a Physical Properties Measurement System (PPMS, Quantum Design, 16T) equipped with a vibrating sample magnetometer (VSM) option and a Magnetic Properties Measurement System (MPMS, Quantum Design,7T). The AC magnetic susceptibility was measured at various frequencies in the MPMS under an excitation field H = 10Oe. Due to the minimal in-plane anisotropy of magnetization [Figure S1], we no longer specify the precise magnetic field direction within the ab plane.

The heat capacity of $Ce_3ScBi_5$ was measured by the thermal relaxation method in the PPMS, with samples securely attached to the platform using Apiezon N grease. Electronic resistivity measurements of $Ce_3ScBi_5$ were conducted on the PPMS with a four-point collinear geometry. A DC of 3000 μA was applied to measure the resistivity under zero-field conditions.

Table S1. Crystallographic and structure refinement data of $Ce_3ScBi_5$.

| Chemical formula | **$Ce_3ScBi_5$** |
|---|---|
| Temperature | 274(2) K |
| Formula weight | 1510.23 g/mol |
| Radiation | Mo $K\alpha$ 0.71073 Å |
| Crystal system | Hexagonal |
| Space group | $P6_3/mcm$ |
| Unit-cell dimensions | $a$ = 9.6628(9) Å |
| | $b$ = 9.6628(9) Å |
| | $c$ = 6.5181(9) Å |
| Volume | 527.06(12) Å$^3$ |
| Z | 2 |
| Density (calculated) | 9.516 g/cm$^3$ |
| Absorption coefficient | 96.332 mm$^{-1}$ |
| F(000) | 1220 |
| $\Theta$ range for data collection | 2.43 to 34.91° |
| Index ranges | -12<=h<=15, |
| | -15<=k<=15, |
| | -10<=l<=10 |
| Independent reflections | 452 [$R_{(int)}$ = 0.0873] |
| Structure solution program | SHELXT 2018/2 (Sheldrick, 2018) |
| Refinement method | Full-matrix least-squares on F$^2$ |
| Refinement program | SHELXL-2018/3 (Sheldrick, 2018) |
| Function minimized | $\Sigma$ w(F$_o^2$ - F$_c^2$)$^2$ |
| Data/restraints/parameters | 452 / 0 / 42 |
| Goodness-of-fit on F$^2$ | 1.176 |
| Final R indices | 439 data; I>2σ(I) |
| | R1 = 0.0210, wR2 = 0.0430 |
| | all data |
| | R1 = 0.0221, wR2 = 0.0433 |
| Weighting scheme | w=1/[σ$^2$(F$_o^2$)+(0.0061P)$^2$+9.3712P] |
| | where P=(F$_o^2$+2F$_c^2$)/3 |

Table S2. Atomic coordinates and equivalent isotropic thermal parameters of $Ce_3ScBi_5$.

| Atom | Wyckoff | Symmetry | x | y | z | Occup[a] | $U_{eq}$[b] |
|---|---|---|---|---|---|---|---|
| Ce | 6g | m2m | 0.61773 | 0.61773 | 0.75000 | 1.000 | 0.011 |
| Sc | 2b | -3. m | 1.00000 | 1.00000 | 1.00000 | 1.000 | 0.010 |
| Bi1 | 4d | 3. 2 | 0.66667 | 0.33333 | 0.50000 | 1.000 | 0.010 |
| Bi2 | 6g | m2m | 1.00000 | 0.73773 | 0.75000 | 1.000 | 0.010 |

[a] *Occup*: Occupancy.
[b] $U_{eq}$: equivalent isotropic thermal parameter.

Table S3. Anisotropic atomic displacement parameters ($Å^2$) for $Ce_3ScBi_5$.

| Lable | $U_{11}$ | $U_{22}$ | $U_{33}$ | $U_{12}$ | $U_{13}$ | $U_{23}$ |
|---|---|---|---|---|---|---|
| Bi(01) | 0.01021(14) | 0.01021(14) | 0.00923(19) | 0 | 0 | 0.00511(7) |
| Bi(02) | 0.01045(17) | 0.00915(14) | 0.01030(17) | 0 | 0 | 0.00523(9) |
| Ce(03) | 0.01026(18) | 0.01026(18) | 0.0101(2) | 0 | 0 | 0.0042(2) |
| Sc(04) | 0.0111(8) | 0.011(8) | 0.0073(13) | 0 | 0 | 0.0055(4) |

The anisotropic atomic displacement factor exponent takes the form: $-2\pi^2[ h^2 a^{*2} U_{11} + ... + 2hka^*b^*U_{12} ]$

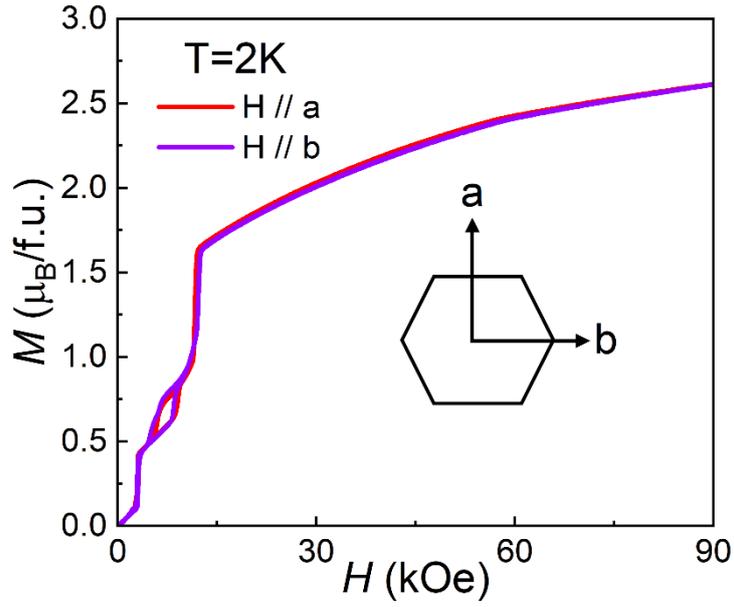

Figure. S1. Isothermal magnetization of $Ce_3ScBi_5$ measured at 2K in ab plane.

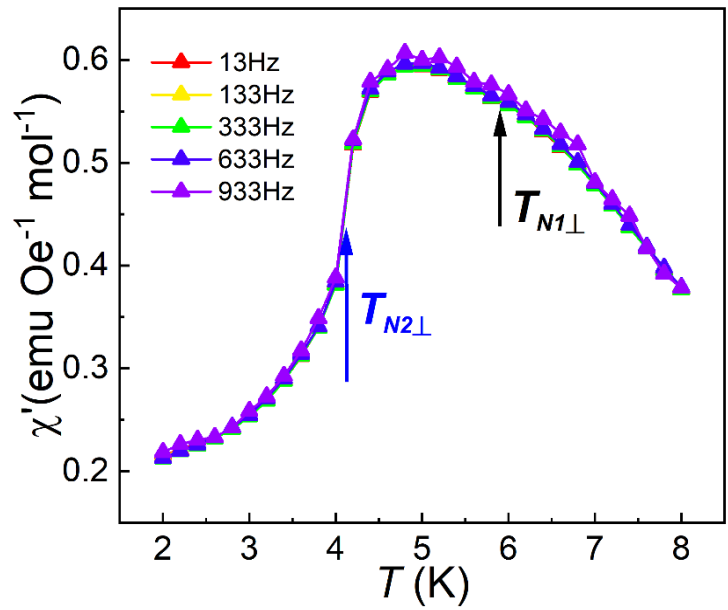

Figure. S2. The ac magnetic susceptibility of $Ce_3ScBi_5$ measured in a fixed magnetic field of 10 Oe. The ac magnetic field was applied perpendicular to the c-axis.

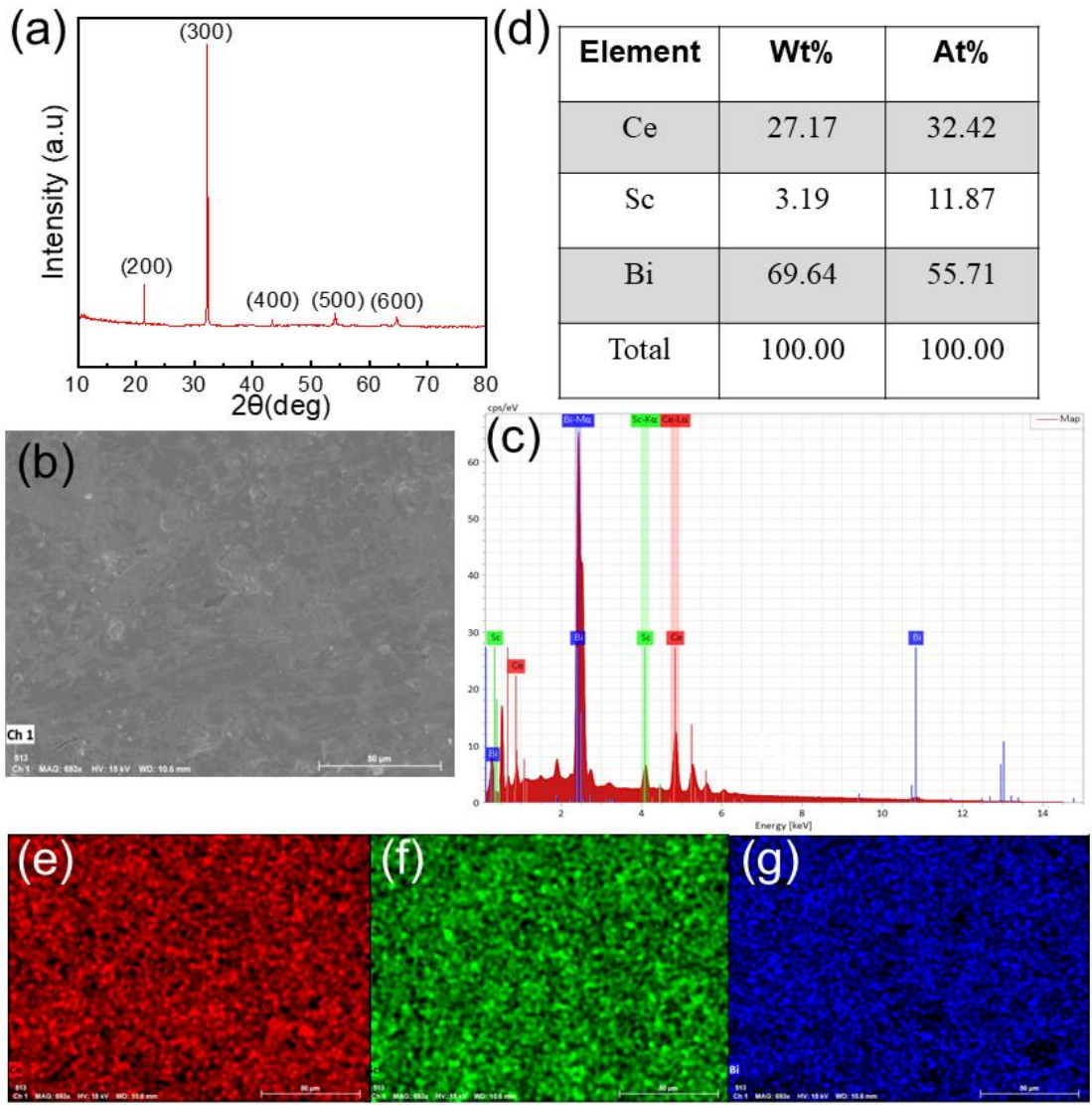

Figure. S3. Characterization of $Ce_3ScBi_5$. (a) The single-face XRD pattern of the (h00) face. SEM image for single crystal in (b) with corresponding energy-dispersive X-ray (EDX) spectra in (c). (d) weight and atomic percentage of Ce, Sc, and Bi atoms. (e), (f), and (g) show EDX elemental color mapping for Ce, Sc, and Bi for the area in (b), respectively.

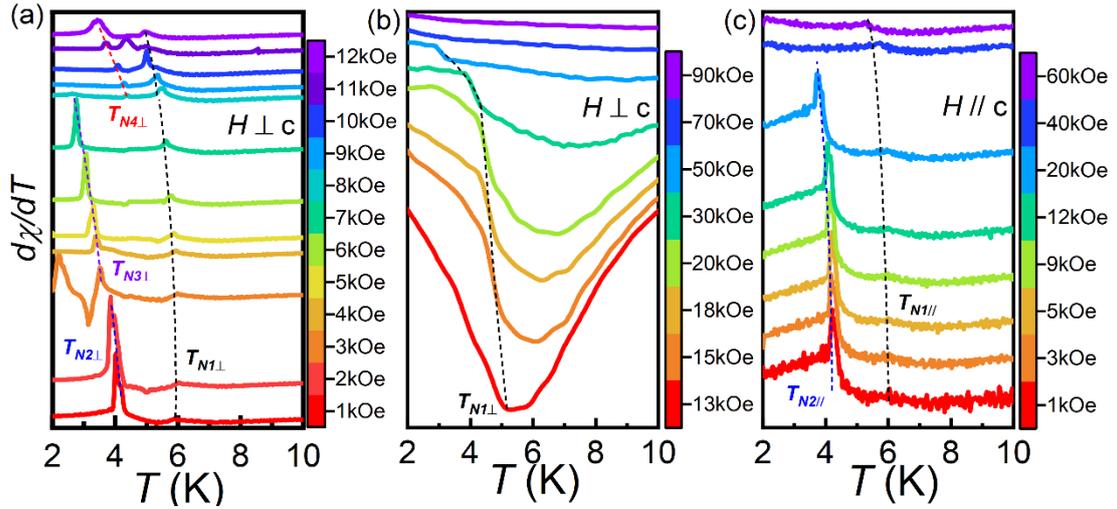

Figure. S4. The derivative of magnetic susceptibility ($d\chi/dT$) for $Ce_3ScBi_5$ under various magnetic fields with the applied field both perpendicular (a), (b), and parallel (c) to the c axis. Data have been shifted vertically for clarity, and the short dot lines are guided to the eye.

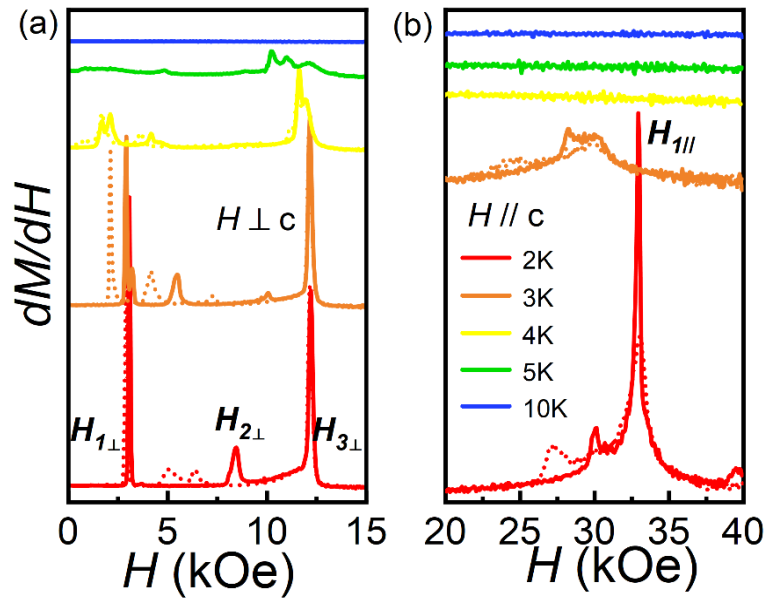

Figure. S5. The $dM/dH$ curves for $Ce_3ScBi_5$ were measured at different temperatures with perpendicular (a) and parallel (b) to the c axis. The solid and dashed lines represent field-sweep up and down measurements respectively.

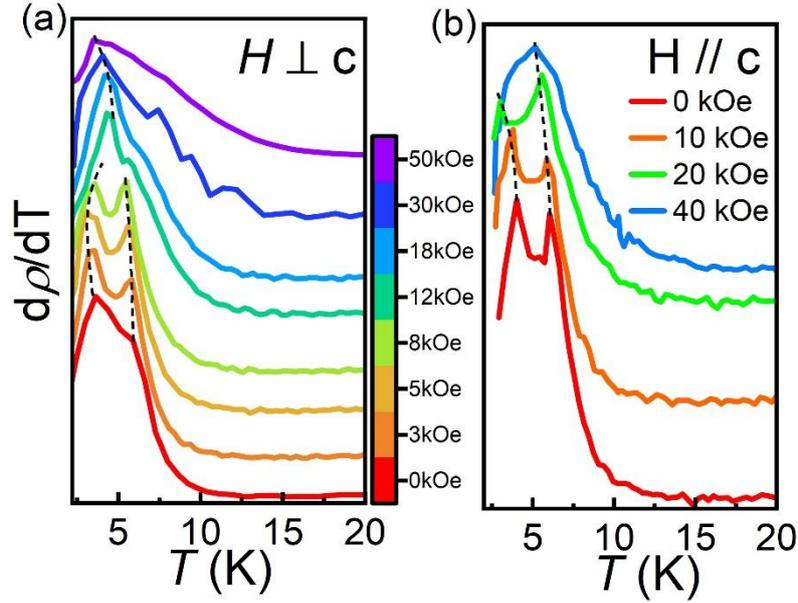

Figure. S6. The derivative of resistivity ($d\rho/dT$) for $Ce_3ScBi_5$ under various magnetic fields with the applied field both perpendicular (a) and parallel (b) to the c axis.

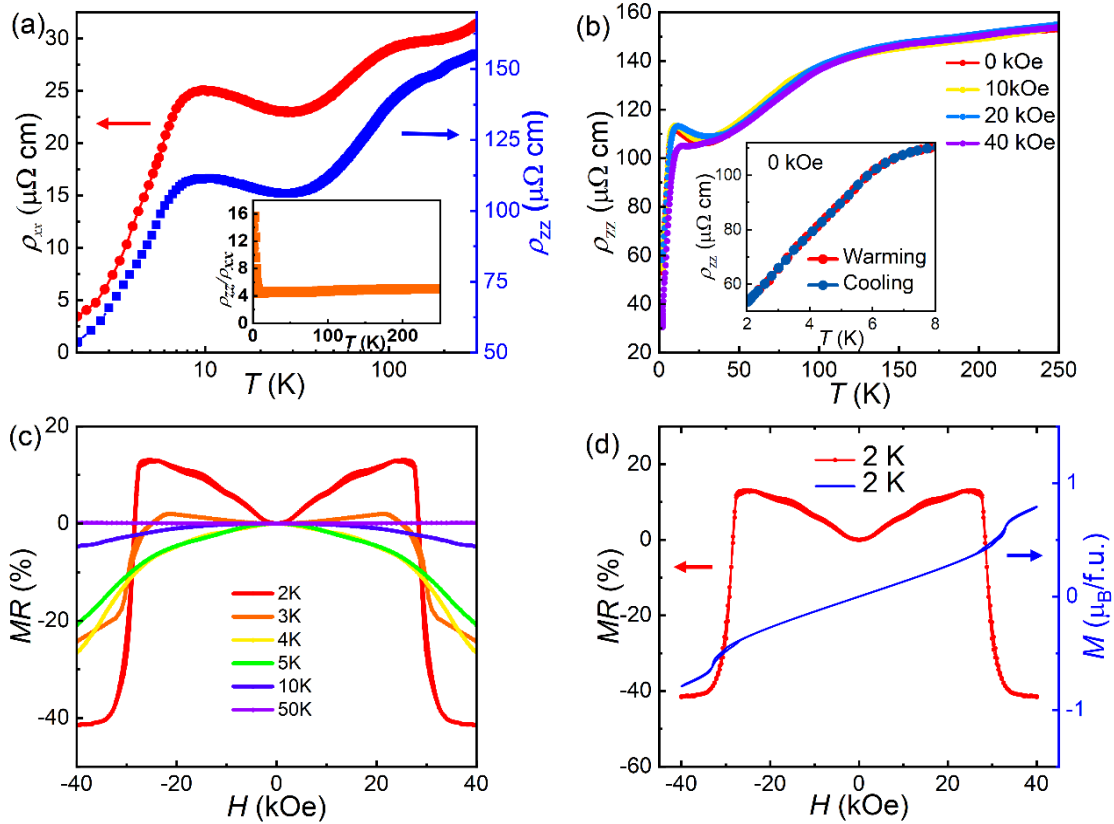

Figure S7. (a) The $\rho_{xx}(T)$ ($I // [100]$) and $\rho_{zz}(T)$ ($I // [100]$) curves of $Ce_3ScBi_5$. The inset shows the radio of $\rho_{zz}/\rho_{xx}$ (b) Temperature dependence of $\rho_{zz}(T)$ for $H // c$ under various magnetic fields. The inset shows no thermal hysteresis from the transitions. (c) Isothermal magnetoresistance for $H // c$ at various temperatures. (d) Field-dependent $MR$ (red) and $MH$ (blue) curves at 2 K for $H // c$.